\documentclass{article}

\usepackage{setspace}
\usepackage[final,nonatbib]{neurips_2023}
\makeatletter
\def\@noticestring{}
\makeatother

\usepackage[utf8]{inputenc}
\usepackage[T1]{fontenc}
\usepackage{xurl}
\usepackage{hyperref}
\usepackage{booktabs}
\usepackage{amsfonts}
\usepackage{nicefrac}
\usepackage{microtype}
\usepackage{graphicx}
\usepackage{tabularx}
\usepackage{longtable}
\usepackage[dvipsnames,table,xcdraw]{xcolor}

\definecolor{light-gray}{HTML}{E5E4E2}
\usepackage[most]{tcolorbox}
\usepackage{enumitem}
\usepackage{arabtex}
\usepackage{utf8}
\hypersetup{
  linkcolor  = violet!85!black,
  citecolor  = blue!30!black,
  urlcolor   = Aquamarine!50!black,
  colorlinks = true,
}
\hypersetup{breaklinks=true}
\usepackage{xurl}
\usepackage[hypcap=false,font={bf}]{caption}

\setlist[enumerate]{leftmargin=*, label=\arabic*.}
\setlist[itemize]{leftmargin=*}
\setlist[itemize,1]{label=\textbullet}
\setlist[itemize,2]{label=\textopenbullet}

\usepackage[backend=biber,
notes,short]{biblatex-chicago}
\AtEveryBibitem{%
\clearlist{language}
\ifentrytype{article}{
\clearfield{urlyear}%
\clearfield{pubstate}%
}{}
}

\usepackage{array,booktabs,calc}
\usepackage{titlesec}
\titleformat{name=\subsubsection}
  {\normalfont\bfseries\itshape}{\thesubsubsection}{2em}{\hspace{-1em}}{}
\titleformat{name=\subsubsection,numberless}
  {\normalfont\bfseries\itshape}{}{0cm}{\hspace{0cm}}{}

\title{Oversight for Frontier AI through a Know-Your-Customer Scheme for Compute Providers}

\author{%
  \parbox{\linewidth}{\centering\bfseries%
  Janet Egan$^{1}$ and Lennart Heim$^{2}$
  \thanks{Correspondence to \texttt{janetegan@hks.harvard.edu} and \texttt{lennart.heim@governance.ai}}}\\
  \parbox{\linewidth}{\centering
  $^{1}$Harvard Kennedy School, $^{2}$Centre for the Governance of AI
  }
}

\begin{document}
\maketitle

\begin{abstract}
To address security and safety risks stemming from highly capable
artificial intelligence (AI) models, we propose that the US government
should ensure compute providers implement Know-Your-Customer (KYC)
schemes. Compute -- the computational power and infrastructure required
to train and run these AI models -- is emerging as a node for oversight.
KYC, a standard developed by the banking sector to identify and verify
client identity, could provide a mechanism for greater public oversight
of frontier AI development and close loopholes in existing export
controls. Such a scheme has the potential to identify and warn
stakeholders of potentially problematic and/or sudden advancements in AI
capabilities, build government capacity for AI regulation, and allow for
the development and implementation of more nuanced and targeted export
controls. Unlike the strategy of limiting access to AI chip purchases,
regulating the \emph{digital access} to compute offers more precise
controls, allowing regulatory control over compute quantities, as well
as the flexibility to suspend access at any time. To enact a KYC scheme,
the US government will need to work closely with industry to (1)
establish a dynamic threshold of compute that effectively captures
high-risk frontier model development, while minimizing imposition on
developers not engaged in frontier AI; (2) set clear requirements and
guidance for compute providers to keep records and report high-risk
entities; (3) establish government capacity that allows for co-design,
implementation, administration and enforcement of the scheme; and (4)
engage internationally to promote international alignment with the
scheme and support its long-term efficacy. While the scheme will not
address all AI risks, it complements existing proposed solutions by
allowing for a more precise and flexible approach to controlling the
development of frontier AI models and unwanted AI proliferation.
\end{abstract}

\clearpage
\addtocounter{footnote}{-1}

\section*{Executive Summary}\label{executive-summary}

Emerging risks associated with the development of frontier AI
models\footnote{Defined as `highly capable foundation models that could
  exhibit dangerous capabilities', \cite{anderljung_frontier_2023}} warrant additional regulatory intervention by the US
government. The potential of these AI capabilities to enhance
adversarial military capabilities and facilitate human rights abuses has
led the US to introduce export controls that restrict exports of the
specialized AI chips required to develop and deploy large AI models,
among other restrictions. Yet gaps in these controls have emerged: there
are currently no restrictions on entities accessing controlled chips and
their associated computing power through digital means, such as cloud
compute provision, offering a potential avenue for adversarial
militaries and non-state actors of concern to benefit from US
technology. While a blanket ban on cloud access could harm US technology
leadership and would be difficult to enforce, there are clear security
grounds for addressing these proliferation risks.

At the same time, broader risks to security and public safety are
eliciting concern and a willingness to act from industry and government
alike. Experts in industry and academia are warning of significant
misuse risks, such as AI increasing the availability of biological
weapon information and incentivizing malicious actors to attempt their
development,\autocite{zakrzewski_tech_2023} as
well as increasing risks of misinformation and electoral interference.
US AI leaders have committed to following voluntary
guidelines,\footnote{\cite{kang_show_2023}} but as noted by Senator
Blumenthal, Chair of the Senate Judiciary Subcommittee on Privacy,
Technology, and the Law, `there will be regulation. The only question is
how soon, and what.'\autocite{noauthor_oversight_2023} As the home to
leading players in the AI industry and supply chain, the US is uniquely
positioned to shape regulatory approaches. Yet interventions will need
to carefully balance maintaining US influence and industry power with
providing avenues to effectively identify and mitigate critical risks.

Across these proliferation, safety, and security risks, compute -- the
computational power and infrastructure required to train and run these
AI models -- offers a key node for oversight and control. The quantity
of compute required for frontier AI models has resulted in cloud compute
forming a key part of the AI supply chain, with the US as the global
leader in AI compute provision. Alongside other potential regulatory
interventions,\footnote{Such as interventions at the model and
  applications levels, as proposed by Microsoft. \cite{smith_developing_2023}.} increasing oversight of AI compute could
enable earlier identification of emerging risks and more targeted
responses.

This paper recommends that the US government implement a
Know-Your-Customer (KYC) scheme for AI compute providers, most notably
Cloud Service Providers (CSPs), to enable greater oversight of the
development of frontier AI models. Such a concept has already been
proposed by Microsoft,\autocite{smith_how_2023} as
well as AI researchers,\autocite{fist_chinese_2023} as a way of increasing
accountability and managing risks. Implemented in partnership with
industry, a KYC scheme has the potential to warn of significant advances
in AI capability, build government capacity in AI regulation, and allow
for more nuanced and targeted controls. This scheme could be accompanied by updated Export Administration Regulations that restrict the provision of above-threshold compute to companies on the Entity List. Beyond export controls,
a KYC scheme could provide the groundwork for domestic safety
regulations and support responsible AI development. The KYC scheme could
be designed to leverage existing technical metrics and preserve privacy
for compute providers and customers. This paper draws on lessons learned
from the mature application of KYC in the financial sector to propose
the development of a KYC scheme for compute providers. It recommends
that the US government work with industry to:

\begin{enumerate}
\item \textbf{Establish a threshold of compute for the scheme} that
  effectively captures high-risk frontier model development,\footnote{In
    this paper, we are primarily addressing the governance of AI systems
    development. Questions about potential large-scale deployment, i.e.
    inference compute, are outside the scope of this particular
    discussion. However, the oversight of deployment compute might also
    be a future policy tool for regulating AI deployment (Appendix A in \cite{obrien2023}; Brundage et al., forthcoming).} while minimizing imposition
  on developers not engaged in frontier AI. The threshold should be
  defined by the total amount of computational operations -- a metric
  easily accessible to compute providers, as they employ chip-hours for
  client billing, convertible to total computational operations.
  Additionally, this threshold would need to be dynamic and subject to
  periodic reassessments by government, in close consultation with
  industry, to remain in step with developments in training efficiency
  as well as broader societal changes. It would also need to be
  supported by collaboration between compute providers, as well as with
  government, to minimize evasion risks.
\item \textbf{Set clear requirements for compute providers,} including
  requirements for gathering information, implementing fraud detection,
  keeping records, and reporting to government any entities that match
  government-specified `high-risk' profiles. These requirements should
  be technically feasible, resilient against efforts to evade detection
  and enforceable, while preserving privacy.
\item \textbf{Establish government capacity} within the US Department of
  Commerce that allows for the co-design, implementation,
  administration, and enforcement of the scheme. This capacity should
  draw on existing expertise within the US government, as well as
  contribute to a deeper understanding of AI regulatory challenges to
  inform broader policies.
\item \textbf{Engage with international partners} to promote alignment with
  the scheme. While the US, as a significant global compute provider
  that wields substantial influence in the semiconductor supply chain,
  can exert broad influence through a domestically implemented scheme,
  cooperation with international partners will be a key enabler of
  increased oversight in the longer term. Consistent international
  standards will help ameliorate the risk of diminishing US AI
  leadership and will be essential to the long-term effectiveness of the
  scheme.
\end{enumerate}

In support of this scheme, this paper makes several further
recommendations to the US government, including to engage industry to
co-design the scheme; develop more targeted controls for the cloud;
publish guidance on information sharing in the context of US antitrust
laws to enable effective risk management by CSPs; and strong
international advocacy and engagement to garner international buy-in and
alignment.

\clearpage

\section{Policy Context}\label{policy-context}

\subsection*{Gaps have emerged in the US's AI regulatory
architecture}\label{gaps-have-emerged-in-the-uss-ai-regulatory-architecture}

The US's October 2022 export controls, further strengthened in 2023,\autocite{bureau_of_industry_and_security_commerce_2023} are designed to limit the PRC's
ability to access and develop semiconductors for highly capable
artificial intelligence (AI) applications.\autocite{bureau_of_industry_and_security_commerce_2022} Experts expect this measure to slow the PRC's
development of advanced AI,\autocite{allen_clues_2023} forming a cornerstone of US efforts to counter the
proliferation of US-enabled AI technology to adversarial militaries. A
subsequent executive order in August 2023 complements this, by
inhibiting US investment in China's advanced technology
sector.\autocite{the_white_house_executive_2023} Yet there is evidence of other
avenues through which PRC entities can still access and develop
high-risk advanced AI capabilities. It is not necessary for these
entities to own the physical chips if they are accessible digitally. PRC
actors can use the computing power of controlled US chips through the
cloud, without any formal mechanism for oversight. There are clear
national security grounds for addressing this gap. Regardless of whether
management is achieved through a blanket ban (not recommended, see Box 1
for details) or active monitoring and more targeted controls
(recommended), increased visibility of advanced AI cloud compute is an
essential prerequisite to effective action.

\begin{tcolorbox}[breakable,boxrule=1pt,enhanced jigsaw, sharp corners,pad at break*=1mm,colbacktitle=gray!05,colback=gray!05,colframe=black,coltitle=black,toptitle=1mm,bottomtitle=1mm,fonttitle=\bfseries\centering,parbox=false,title=Box 1: Controlling the PRC's access to chips through the cloud]
As the PRC's access to advanced AI chips through the cloud seems to be
at odds with current export controls, US lawmakers are already looking
to take action. A Bill introduced into Congress in mid July 2023
proposes to `prohibit support for the remote use or cloud use of
integrated circuits' covered by export controls to prevent their use by
actors in China or Macau.\footnotemark\ Yet
a blanket ban on this access would likely lead to adverse
outcomes.\footnotemark\ In particular, it risks diminishing US market share of
cloud computing services and the US hardware used in these services,
while the PRC focuses completely on its own sovereign capability or
establishing access elsewhere.\footnotemark

Unlike physical exports, cloud access allows for a precise and flexible
mechanism to manage proliferation risks. Export controls on chips are
blunt by necessity -- once exported, chips cannot be retracted nor their
end use controlled. So, while a PRC entity's use of a small number of
high-capability chips would not in themselves raise strategic AI
proliferation concerns,\footnotemark\ individual chips exported to the PRC could
be amalgamated to bolster and advance in-country capability. The use of
this capability is then beyond US awareness and control. In contrast,
digital access to these chips only provides entities with point-in-time
compute power, which can be restricted or shut off at any stage. Compute
providers also retain insight into how much compute is being used per
customer. Cloud access therefore allows for flexible digital controls
that can be precisely targeted to capture the most resource-intensive
compute uses, and be easily adapted as geopolitical conditions,
capabilities, and risks change. As the compute provider already has
compute usage information for its customers, this approach does not
require privacy trade-offs to be effective. This allows for a targeted
approach that directly addresses risk, while minimizing impact on US
industry and technology leadership.

However, controlling compute via providers is not a perfect instrument
and we want to be conscious of its limitations. We discuss them in the Appendix below.
\end{tcolorbox}
\addtocounter{footnote}{-3}
\footnotetext{\cite{rep_jackson_closing_2023}.}\stepcounter{footnote}
\footnotetext{\cite{dohmen_controlling_2023}.}\stepcounter{footnote}
\footnotetext{\cite{fist_chinese_2023}.}\stepcounter{footnote}
\footnotetext{China's use of a single, or small
  number of export-controlled chips is unlikely to be of concern to US
  interests, as it would not offer advantage over existing domestic
  Chinese AI capabilities.}

\subsection*{In addition to closing export control gaps, greater
oversight over AI compute can yield broader domestic and international
benefits}\label{in-addition-to-closing-export-control-gaps-greater-oversight-over-ai-compute-can-yield-broader-domestic-and-international-benefits}

In light of increasing concerns around AI misuse and risks to public
safety, governments, industry,\autocite{smith_how_2023} and researchers\autocite{anderljung_frontier_2023} are calling for more oversight and regulation of AI. As
posited by Senator Richard Blumenthal, Chair of the Senate Judiciary
Subcommittee on Privacy, Technology, and the Law (the Subcommittee),
`Congress failed to meet the moment on social media. Now we have the
obligation to do it on AI before the threats and the risks become
real.'\autocite{blumenthal_blumenthal_2023} Congress continues to engage industry on potential AI
regulation. Seven leading US AI companies have agreed to voluntary
ethical, security, and safety standards and called for greater public
oversight, but these measures are not currently enforceable and will be
unlikely to affect the behavior of broader industry players.\autocite{shear_pressured_2023} Microsoft's Brad Smith has called for
KYC-inspired techniques to be used to increase accountability in AI
development.\autocite{smith_developing_2023} As a
key enabler of AI development, compute is emerging as a key domestic and
international governance instrument that can offer greater government
oversight. In particular:

\begin{itemize}
\item It is a useful node to identify entities conducting cutting-edge AI development\footnote{Furthermore, these entities
    could also encompass AI agents themselves. Consider, for instance,
    an AI model exhibiting behaviors akin to a computer worm, actively
    seeking to acquire additional compute resources (e..g, for
    self-replication). Measures should be in place to ensure that such
    entities accessing compute are legitimate legal entities, thereby
    preventing rogue AI systems from taking unauthorized actions.}
  (i.e. those using the greatest amount of training compute), which
  would support the US government in identifying and engaging specific
  industry players on evaluating and managing risks.\autocite{whittlestone_response_nodate}
\item
  Greater oversight would also increase AI exposure and knowledge in the
  US government, enhancing its ability to effectively regulate this
  emerging technology.\autocite{anderljung_frontier_2023} It would provide the US government with a mechanism
  to verify the adherence of the leading AI companies' implementation of
  their voluntary commitments.\autocite{the_white_house_fact_2023} In fact, the cybersecurity provisions in these commitments
  would also need to be carried out among compute providers, in addition
  to at the level of AI developers, in order to be effective. Oversight
  at the compute level would also allow for implementation of potential
  future regulations. For example, potential regulations on frontier AI
  systems might be triggered by such systems passing a predefined
  compute threshold, for which the compute provider can act as a natural
  checkpoint.
\item
  It would also provide the US additional leverage to promote and/or
  enforce AI standards and regulations on an international scale.
  Mustafa Suleyman, CEO of AI start-up Inflection and co-founder of
  DeepMind, has advocated for the US to leverage its chokehold on the
  chip supply chain to shape global AI standards by limiting the
  consumption of cutting-edge US chips to entities that agree to adhere
  to minimum standards.\autocite{waters_us_2023}
\end{itemize}

\subsection*{Cloud compute is an essential enabler of AI
development}\label{cloud-compute-is-an-essential-enabler-of-ai-development}

Compute refers to the computational power and infrastructure that is
required to train and run AI models and systems. Training compute has
grown by a factor of 55 million since 2010, outpacing hardware price
performance.\autocite{sevilla_compute_2022} The intensity of compute required for AI makes
it impractical for most AI designers and operators to build and maintain
their own data processing centers. This has led to compute emerging as a
key node in the AI supply chain, with AI compute providers, most notably
CSPs, renting out the hardware infrastructure, including access to
advanced AI chips. It is in the US's interests that AI innovators
continue to use cloud compute (rather than procuring and maintaining
their own costly hardware capabilities) given that the majority of CSPs
are in the US.\autocite{taylor_data_nodate} Allowing easy access to and use of superior US capabilities
can also make AI innovators more invested in and reliant on leading US
software stacks, like NVIDIA's, reducing the incentives for them to
develop their own with less US oversight. Continuing to enable and
support a strong cloud compute industry in the US will support US
technology leadership more broadly. Notably, US companies are subject to
US regulations, which serve as a useful channel for interventions to
manage AI risks.

\subsection*{Increasing risks call for increasing
vigilance}\label{increasing-risks-call-for-increasing-vigilance}

There is currently no obligation for US cloud providers to actively
verify the identity of their customers before providing them with access
to advanced AI compute. This is problematic in terms of US export
controls. It is also of concern from a broader public interest
perspective. As AI models grow in capability, so too do the risks to
public safety. Foundation models, also known as general-purpose models
(like ChatGPT) have already demonstrated significant capability
advances. AI experts across sectors are warning of risks that may arise
in future generations of general purpose models, particularly because it
is hard to predict how dangerous capabilities might arise, and/or how
they might be misused.\autocite{anderljung_frontier_2023} Researchers have coined the term `frontier AI models',
defined as `highly capable foundation models that could exhibit
dangerous capabilities'.\autocite{anderljung_frontier_2023} For example,
frontier AI models could have the potential to conduct sophisticated,
automated offensive cyber activities, enable non-experts to design and
access new bioweapons, generate and proliferate persuasive deepfake
content, or act in unexpected ways that cause harm.\autocite{shevlane_model_2023} Like their forerunners
ChatGPT, Bard, and other current foundation models, these future
frontier AI models would need to be trained through significant use of
computing power: typically the more computation, the greater the
capability. Therefore, in considering options to prevent technology
transfer to adversarial militaries, the US government should consider
interventions that can also address broader safety and security needs.

\begin{tcolorbox}[boxrule=1pt,enhanced jigsaw, sharp corners,pad at break*=1mm,colbacktitle=gray!05,colback=gray!05,colframe=black,coltitle=black,toptitle=1mm,bottomtitle=1mm,fonttitle=\bfseries\centering,parbox=false,title=Box 2: What do we mean by `compute providers'?]
In this context, we use the term `compute provider' to refer to an
entity providing access to computing power and infrastructure. In the
majority of cases for AI development, these compute providers are CSPs,
businesses that provide access through the cloud (including to foreign
entities). Yet given the potential domestic safety implications of
frontier AI models, taking a more expansive view of `compute providers'
may increase the longevity and impact of the proposed scheme. From a
domestic perspective, it leaves the scheme open to the possibility of
capturing providers of compute hardware and in-house compute capability
(when it has the ability to go above the designated threshold). For
example, it could be desirable that large technology companies using
their AI compute internally also monitor and keep records for compliance
with regulations in alignment with a KYC scheme. This is an area that
would benefit from further research and analysis in collaboration with
compute providers.
\end{tcolorbox}

\section{A Know-Your-Customer scheme for compute
providers}\label{a-know-your-customer-scheme-for-compute-providers}

The US government should introduce a KYC scheme that ensures adequate
monitoring for advanced AI cloud compute and allows the government to
require compute providers to report high-risk entities and deny access
to entities of concern.\autocite{fist_chinese_2023} A KYC scheme requires businesses
and organizations to verify the identity of their clients in order to
provide them with access to particular goods and services. Introducing
KYC requirements for entities accessing significant amounts of compute
could help identify risks and enable further targeted restrictions where
there is significant risk to national and global interests. Requiring
compute providers to build greater awareness of the risks could
encourage a safer AI industry aligned with public benefit.

It is important to note that this proposed KYC scheme for advanced AI
cloud compute would not capture all AI models being developed or used by
malicious actors. For example, there are already specific less-advanced
models today, which do not require massive amounts of compute, that
raise biochemical weapon development concerns\autocite{urbina_dual_2022} or
enable more targeted malicious cyber activity. Setting the threshold to
capture and monitor the compute of all AI models would not be
beneficial, as it would capture too much information to be useful while
imposing a significant imposition on industry. Such risks could instead
be managed through other safeguards,\footnote{For example, through
  Government-Industry engagement on malicious cyber activity,
  intelligence monitoring of actors of concern etc.} while the proposed
in-depth KYC would focus on powerful foundation models trained on
significant amounts of compute.

The history of the implementation of KYC in the financial sector could
provide useful lessons for scheme design (Box 3).

\begin{tcolorbox}[boxrule=1pt,enhanced jigsaw, sharp corners,pad at break*=1mm,colbacktitle=gray!05,colback=gray!05,colframe=black,coltitle=black,boxrule=0pt,toprule=1pt,leftrule=1pt,bottomrule=0pt, titlerule=1pt,rightrule=1pt,toptitle=1mm,bottomtitle=1mm,fonttitle=\bfseries\centering,parbox=false,title=Box 3: Learning from KYC in the financial sector]
The most mature application of KYC can be seen in the financial sector,
where it forms the cornerstone of Anti-Money Laundering (AML) and
Counter-Terrorist Financing (CTF) obligations. In this context, KYC is
implemented through domestic US legislation (particularly through the
\href{https://www.occ.treas.gov/topics/supervision-and-examination/bsa/index-bsa.html}{\emph{Bank
Secrecy Act}} including as amended by the
\href{https://www.fincen.gov/resources/statutes-regulations/usa-patriot-act}{\emph{USA
PATRIOT Act}}). This legislation creates obligations for financial
institutions to:\vspace{-3pt}
\begin{itemize}
\item 
  implement a customer identification program
\item conduct risk assessments
\item undertake enhanced due diligence for customers assessed as higher risk
\item identify and verify a customer's beneficial owners
\item conduct ongoing monitoring
\item report suspicious activity to the US government.\footnotemark
\end{itemize}

These obligations are enforced by the
\href{https://www.fincen.gov/}{Financial Crimes Enforcement
Network} (FinCEN) within the US Treasury, which acts as a regulator,
investigator and financial intelligence unit.\footnotemark\ In recent years, FinCEN has
faced enforcement difficulties from the use of shell companies to hide
illicit activity.\footnotemark\ In
response, FinCEN has implemented a rule establishing a beneficial
ownership information reporting requirement, which will obligate
companies registered in the US to provide information on the persons who
control them, with a start date of January 2024.\footnotemark\

\end{tcolorbox}
\addtocounter{footnote}{-4}

\footnotetext{\cite{office_of_the_comptroller_of_the_currency_bank_2019,financial_crimes_enforcement_network_usa_nodate}.}\stepcounter{footnote}

\footnotetext{\cite{financial_crimes_enforcement_network_usa_nodate}.}\stepcounter{footnote}
\footnotetext{\cite{hamilton_us_2021}.}\stepcounter{footnote}

\newpage

\footnotetext{\cite{financial_crimes_enforcement_network_what_nodate}.}\stepcounter{footnote}

\begin{tcolorbox}[boxrule=1pt,enhanced jigsaw, sharp corners,pad at break*=1mm,colbacktitle=gray!05,colback=gray!05,colframe=black,coltitle=black,boxrule=0pt,toprule=0pt,leftrule=1pt,bottomrule=1pt,rightrule=1pt, titlerule=1pt,toptitle=1mm,bottomtitle=1mm,fonttitle=\bfseries\centering,parbox=false]

Given the global nature of financial markets and transactions, KYC for
the financial sector is well supported by international cooperation. In
particular, the
\href{https://www.fatf-gafi.org/en/the-fatf/what-we-do.html}{Financial
Action Task Force} (FATF) is the key intergovernmental organization
that promotes global AML/CTF standards.\footnotemark\ The FATF was established by the Group of 7
(G7) in 1989.\footnotemark\ The FATF standards
include 40 recommendations -- including on KYC requirements -- that form
a framework for member nations to implement through their own domestic
legislation.\footnotemark\ The FATF also publishes lists of jurisdictions
that fail to effectively implement safeguards, and calls for members to
apply enhanced due diligence and countermeasures to entities from these
countries to protect the international financial system.\footnotemark\ This ensures a consistent approach that supports efforts to
counter money laundering, terrorism financing, and the proliferation of
weapons of mass destruction.\footnotemark\
The financial sector's KYC scheme is also supported by a
well-established intelligence architecture specifically for AML/CTF
threats. The \href{https://egmontgroup.org/about/}{Egmont Group
of Financial Intelligence Units} brings together 166 Financial
Intelligence Units globally to share intelligence and expertise to
counter money laundering and terrorism financing threats.\footnotemark\

The cost of compliance and enforcement of AML/CTF is significant. The
2024 Federal Budget allocated \$229 million for FinCEN, including
funding for 350 staff members, an increase of \$39 million from
2023.\footnotemark\ In 2022, the cost of financial crime
compliance efforts across financial institutions is estimated to be
\$40.7 billion in the US and \$274.1 billion globally.\footnotemark\
 
\emph{Implementation: obstacles and adjustments}

The financial sector presents a case study of situations in which
non-compliance with obligations persisted until significant penalties
were applied. For the first decade of the Bank Secrecy Act, from 1972 to
1985, regulators did not enforce reporting requirements, resulting in
low compliance from financial institutions.\footnotemark\ However, a 1985 \$500,000 penalty issued to the Bank
of Boston led to a sharp increase in reporting, as well as more evasive
behavior from customers.\footnotemark\ More recently, Congress has
taken further action to increase enforcement. The Anti-Money Laundering
Whistleblower Improvement Act, signed into law in December 2022,
increases reporting incentives with greater financial rewards for
successful tips.\footnotemark

KYC in the financial sector also demonstrates implementation risks,
including the use of ``structuring'' to evade publicly set
thresholds.\footnotemark\ In response to obligations
for banks to report transactions exceeding \$10,000 in any one day,
entities and individuals started to intentionally break up transactions
across bank accounts and/or days to avoid scrutiny.\footnotemark\ Amendments made through the \emph{2001} \emph{PATRIOT
Act} have sought to address this by making structuring a criminal
offense.
\end{tcolorbox}

\addtocounter{footnote}{-11}

\footnotetext{\cite{financial_action_task_force_what_nodate}.}\stepcounter{footnote}
\footnotetext{\cite{international_monetary_fund_anti-money_nodate}.}\stepcounter{footnote}
\footnotetext{\cite{financial_action_task_force_fatf_nodate}.}\stepcounter{footnote}
\footnotetext{\cite{financial_action_task_force_high-risk_2023}.}\stepcounter{footnote}
\footnotetext{\cite{lattin_australia_2023}.}\stepcounter{footnote}
\footnotetext{\cite{egmont_group_about_2023}.}\stepcounter{footnote}
\footnotetext{\cite{office_of_management_and_budget_budget_2023}.}\stepcounter{footnote}
\footnotetext{\cite{lexisnexis_risk_solutions_true_nodate}.}
\footnotetext{\cite{linn_redefining_2010}.}\stepcounter{footnote}
\footnotetext{\cite{linn_redefining_2010}.}\stepcounter{footnote}
\footnotetext{\cite{sen_grassley_bill_2022}.}\stepcounter{footnote}
\footnotetext{\cite{linn_redefining_2010}.}\stepcounter{footnote}
\footnotetext{\cite{federal_financial_institutions_examination_council_bsaaml_nodate}.}

\newpage

\subsection*{Developing a KYC scheme for advanced AI cloud
compute}\label{developing-a-kyc-scheme-for-advanced-ai-cloud-compute}

Informed by the model established by the financial sector, the
development of a new KYC scheme will require:
\begin{enumerate}
\item defining a threshold of AI compute at which the scheme would apply
\item introducing requirements and guidance for compute providers above that threshold, including reporting to government entities that match   specified `high-risk' profiles, and adhering to rules
\item establishing government capacity for engagement, regulation, and enforcement
\item engaging and cooperating with international partners
\item establishing a process for evaluation and updates to the scheme.
\end{enumerate}

The introduction of a KYC scheme for advanced AI cloud compute would lay
the groundwork for understanding the threat, including any substantial
access attempts by People's Liberation Army (PLA)-linked entities, and
enable more targeted restrictions to prevent entities of concern from
access through the cloud. Importantly, it would also increase
government's ability to see trends and emerging risks and point AI
policy makers towards companies operating at the cutting-edge to allow
for better engagement in risk management.

\subsection{Compute is indicative of AI capabilities and KYC
thresholds should be set
accordingly}\label{compute-is-indicative-of-ai-capabilities-and-kyc-thresholds-should-be-set-accordingly}

KYC obligations should apply at and beyond a threshold of advanced AI
compute that captures the most critical AI risks, while minimizing
regulatory impost on industry.\footnote{Here we
  focus on compute used for the development of advanced AI systems, and
  potential large scale deployment. We do not answer the question of
  inference compute.} One can quantify the computational performance of
hardware in terms of a unit of floating point operations per second
(FLOP/s). The resulting accumulation of computing power going into
developing or deploying an AI model can be measured in the quantity of
FLOP.\autocite{heim_flop_2023}
This allows for a defined threshold. Government could work closely with
AI experts and compute providers to ensure that the threshold is set at
a level that captures frontier AI models. Because such models are most
likely to emerge at the largest compute scales, the initial threshold
could be set at the level of FLOP used to train current foundational
models, like GPT-4\autocite{mcaleese_retrospective_2023} or Claude 2,\autocite{wiggers_anthropics_2023} thereby also capturing models trained on even more compute.
This would also ensure that the burden of compliance with the KYC scheme
only falls on operators able to absorb it; GPT-4, for example, is
estimated to have cost \$50 million to train.\autocite{epoch_key_2023} This would only capture a
handful of models at the time of writing. With training requirements
continuing to double every six-to-12 months, we can expect that an
increasing number of cloud-trained AI model developers will be subject
to KYC.\autocite{sevilla_compute_2022}

Improvements in algorithmic efficiency over time may result in increased
compute efficiency -- requiring less compute to achieve the same
training results, and potentially necessitating a lower threshold to
capture models of concern.\autocite{anderljung_frontier_2023,tucker_social_2020} Conversely, advances in
regulation or society's ability to manage AI impacts may decrease the
risks associated with advanced AI models, thereby making a higher
threshold appropriate.\autocite{anderljung_frontier_2023,tucker_social_2020} The threshold should
therefore be dynamic and subject to regular review by government and
industry. It should be responsive to metrics broader than computing
power that influence AI capability and society's resilience to AI risks.
This will also allow for continued refinement as processes mature and
the ability to adapt to changing geostrategic conditions and risks.

Applying a specified FLOP threshold offers a feasible path to
implementation and does not require cloud providers to access the data
or confidential information of their customers. Cloud access to chips is
billed by the hour, so the accumulated total FLOP is easily identifiable
by the compute provider. The compute provider could then implement KYC
checks and enhanced due diligence for any projects seeking to cross that
threshold. In many cases, the total amount of compute procured will be
specified at the time of entering into contract, but there may also be
cases where additional compute is purchased over time, to the point at
which a specific vendor crosses the threshold. Compute providers should
therefore continuously monitor compute use, and ensure that entities
approaching the threshold are funneled into the KYC process before that
point is reached.

\subsubsection*{Regulatory impost is likely to be low, with few stakeholders
affected}\label{regulatory-impost-is-likely-to-be-low-with-few-stakeholders-affected}

Given the proposed threshold, only a small number of customers for a
small number of US compute providers would be affected. Providers that
offer, or use in-house, the most advanced computing power also tend to
be the most resourced, such as Microsoft Azure, Google Cloud, NVIDIA,
Amazon Web Services (AWS), and Oracle Cloud Infrastructure.\autocite{nvidia_corporation_nvidia_2023} These factors could help mitigate the risk of an overly
costly, burdensome regime (criticisms often directed at KYC in the
financial sector). In addition to having the bandwidth to implement KYC,
these companies may already be working to control significant AI risks,
given their public commitments to ethical and/or responsible
AI.\autocite{croak_google_2023} Microsoft has
specifically called for the implementation of a KYC program in their
report \emph{Governing AI: A Blueprint for the Future,}\autocite{smith_how_2023} and Amazon, Anthropic, Google,
Inflection, Meta, Microsoft, OpenAI and NVIDIA, among others, have
committed to further safeguards against risky AI.\autocite{shear_pressured_2023}

\subsection{Requirements on compute providers -- due diligence
that identifies risk and implements
controls}\label{requirements-on-compute-providers-due-diligence-that-identifies-risk-and-implements-controls}

For AI cloud compute services that possess sufficient computational
power, i.e., type and number of AI chips, to surpass the threshold, the
scheme would require compute providers to identify and verify the entity
and its beneficial owners, and maintain appropriate records. The
government could also define `high-risk' profiles, and require compute
providers to report such cases, in order to monitor for emerging risks
and to inform future controls. The KYC scheme should also be used as a
key mechanism to ensure the implementation of rules.

While the KYC scheme provides the mechanism for oversight of
advanced AI compute, rules governing US companies in their provision
of such compute can be implemented through complementary existing
authorities. For example, the US government could update the rules affecting the
Export Administration Regulations to prevent the provision of
above-threshold compute to entities on the Entity List
without a license.\autocite{bureau_of_industry_and_security_commerce_2022}~\footnote{ This paper does not propose expanding export controls to prevent entities beyond the Entity List from accessing cloud compute. Instead, we recommend KYC is used to develop a more nuanced understanding of the risks of PRC use of US compute. Given the flexible, digital controls it enables, it can be quickly applied once risks are identified. 
}

Beyond export controls, the KYC scheme will allow for the implementation
of broader regulations on AI. For example, should the government choose
to mandate the voluntary commitments made by leading AI companies --
including, for example, having implemented safe-development practices
and cybersecurity standards -- the KYC checks could also ensure that
those accessing advanced AI compute are in compliance before permitting
access. In this way, the KYC scheme forms a flexible foundation to
ensure oversight of an increasingly sensitive set of technologies.

\begin{figure}[htb]
\vspace{1mm}
    \centering
    \caption{Mechanisms of KYC Scheme for compute providers}    \label{fig:1}

    \vspace{2mm}
    \includegraphics[width=\linewidth]{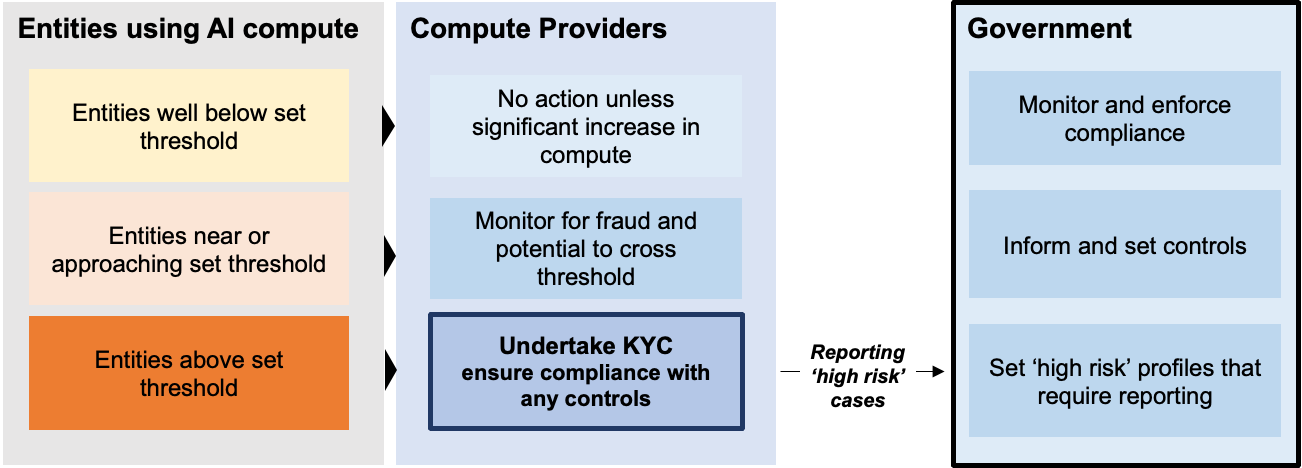}
    \vspace{1mm}
\end{figure}

While we can seek to adapt the model established by the financial sector
to the context of compute providers as a starting point (Box 4),
consultation and stress-testing with industry stakeholders will be key
to developing a workable scheme.

\begin{tcolorbox}[boxrule=1pt,enhanced jigsaw, sharp corners,pad at break*=1mm,colbacktitle=gray!05,colback=gray!05,colframe=black,coltitle=black,boxrule=0pt,toprule=1pt,leftrule=1pt,bottomrule=0pt, titlerule=1pt,rightrule=1pt,toptitle=1mm,bottomtitle=1mm,fonttitle=\bfseries\centering,parbox=false,title=Box 4: Possible requirements for compute providers]
Building on requirements established in the financial sector, and
requiring further consultation with compute providers, KYC requirements
for advanced AI cloud compute might include:\vspace{-5pt}
\begin{itemize}
\item Identifying the entity, including:
\begin{itemize}
    \item company name
    \item proof of incorporation
    \item legal form and status
    \item address of registered office
    \item list of directors and senior management
    \item list of board members
    \item basic regulating powers (e.g. memorandum and articles of association)
    \item unique identifier (e.g. tax identification number or equivalent, where
    applicable).
\end{itemize}
\item
  Identifying key personnel and all beneficial owners,\footnotemark\ including:
    \begin{itemize}
    \item full name, including any aliases
    \item date and place of birth
    \item nationality
    \item home address
    \item government issued identification number.
    \end{itemize}

\item Verifying information provided, including through checking domestic
  and international government registries.
\item Providing a high-level overview of the purpose for the use of compute
  power, and where relevant, investigating details of sublessors of the
  compute.
\end{itemize}

\end{tcolorbox}
\footnotetext{As
    defined in FinCEN's final rule implementing beneficial ownership
    information reporting requirements, a beneficial owner is `any
    individual who, directly or indirectly, either (1) exercises
    substantial control over a reporting company, or (2) owns or
    controls at least 25 percent of the ownership interests of a
    reporting company. \cite{financial_crimes_enforcement_network_beneficial_2022}}
  \vspace*{\fill}    
\newpage

\begin{tcolorbox}[boxrule=1pt,enhanced jigsaw, sharp corners,pad at break*=1mm,colbacktitle=gray!05,colback=gray!05,colframe=black,coltitle=black,boxrule=0pt,toprule=0pt,leftrule=1pt,bottomrule=1pt, titlerule=1pt,rightrule=1pt,toptitle=1mm,bottomtitle=1mm,fonttitle=\bfseries\centering,parbox=false]
\begin{itemize}
  
\item Conducting ongoing high-level usage monitoring to identify changes or emerging characteristics that could change the assessment:
    \begin{itemize}
    \item changes in contracts that bring entities into the KYC threshold or
      increase in procured cloud compute that exceeds the expected scope of
      the stated project
    \item use of compute different from what would be expected from the stated
      purpose (e.g. high-level usage patterns that may indicate AI training
      rather than AI deployment).
    \end{itemize}

\item Sharing information with other compute providers to identify and
  mitigate evasion attempts while preserving privacy.

\item Assessing whether an entity would match a defined `high-risk' profile and reporting these cases to the government. Factors that could be
  considered include:
    \begin{itemize}
    \item the US specially designated nationals and blocked persons
      list;\footnotemark\
    \item the US Entity List.\footnotemark\
    \item other entities restricted from accessing advanced AI chips under
      export controls.\footnotemark\
    \item strong links to a country of concern, which may be informed by factors including that:

      \begin{itemize}
      \item the entity or beneficial owner/s are based in a country of concern
      \item there is evidence that the entity or beneficial owner/s have
        significant ties to a country of concern
      \item the entity's board has significant ties to a country of concern
      \item director/s or senior management are currently affiliated with research institutions from a country of concern
      \item IP addresses originate in a country of concern
      \item evidence of large amounts of data going to/from a country of concern.
      \end{itemize}
    \item the source of the entity's capital or funding is not clear (which
      could potentially lead to requiring the entity to clarify their source
      or lose compute access).
    \item entities from countries on the
      \href{https://www.fatf-gafi.org/content/fatf-gafi/en/publications/High-risk-and-other-monitored-jurisdictions/Call-for-action-June-2023.html}{FATF
      high-risk jurisdiction list}.\footnotemark\
    \item requests for significantly more compute power than is typically used to develop current cutting-edge models. 
    \end{itemize}
\item Using KYC to enforce established rules, which may include:
    \begin{itemize}
    \item updated Export Administration Regulations that restrict companies
      providing above-threshold compute to entities on the
      Entity List
    \item seeking confirmation of the entity's implementation of
      safe-development practices, including cybersecurity standards (if the
      voluntary commitments agreed upon between industry and the White House
      become mandatory).\footnotemark\
    \end{itemize}
\item Maintaining records on the provision or denial of above-threshold
  compute to aid in investigations or demonstrate compliance, when
  needed.
\end{itemize}
\end{tcolorbox}
\addtocounter{footnote}{-4}
\footnotetext{\cite{office_of_foreign_assets_control_specially_2023}.}\stepcounter{footnote}
\footnotetext{\cite{bureau_of_industry_and_security_export_2023}.}\stepcounter{footnote}
\footnotetext{\cite{bureau_of_industry_and_security_commerce_2022}.}\stepcounter{footnote}
\footnotetext{As at August 2023, these are DPRK, Iran and Myanmar. \cite{financial_action_task_force_high-risk_2023}.}\stepcounter{footnote}
\footnotetext{\cite{the_white_house_fact_2023}.}

\subsubsection{Technical feasibility}\label{technical-feasibility}

It is likely that information pertaining to an entity's identity and
beneficial owners will be readily attainable, given that entities would
have such information on hand for their financial institutions.
Similarly, checking information against government registries and lists is also feasible, and can draw on existing compliance expertise from the financial sector.

However, while compute providers can collect statements from customers
on their planned use of cloud compute, this can be difficult to verify in practice. CSPs often take pride in their ability to offer privacy to their customers, with some providers designing `confidential compute' offerings to make it technically impossible for compute providers to look in at the customer's data.\autocite{nvidia_corporation_nvidia_nodate} It is not always clear what compute is
being used for, with both the training of foundational models and using
AI models for inference requiring intensive compute power.\autocite{heim_compute_2021} Given the sensitive
proprietary information and data involved in cutting-edge AI models,
requirements that significantly affect privacy will likely generate
significant industry backlash and diminish US industry power. The
dispute between the FBI and Apple in 2016 is evidence of the tension
between the US government's security priorities and privacy principles
held by the technology sector and general public.\autocite{kharpal_apple_2016} Further research and collaboration with industry is
warranted to identify mechanisms that allow for more effective
verification in a way that preserves privacy. A helpful starting point
could be focusing on the types of clusters used and how the GPUs are
networked, as well as chip hours, which tend to differ according to
purpose. This information is known to the compute provider, as these
requirements would generally be specified as part of a customer order.
Thus, the implementation of the KYC scheme would not require the compute
provider to access the underlying code, data, or any system level
insights, maintaining appropriate privacy standards.

\subsubsection{Mitigating attempts to evade detection}\label{mitigating-attempts-to-evade-detection}

As seen in the case of KYC in the financial sector, malicious actors may
seek to evade detection by engaging in ``structuring''. In the case of
AI, that structuring could involve finding ways to deconstruct
compute-intensive projects into smaller, discrete sub-projects that fall
below the reporting threshold. Compute providers would be responsible
for undertaking their own fraud prevention to ensure that they identify
a single entity acting as multiple customers, a measure that is likely
already present in existing practices to prevent customers from
bypassing terms and conditions. The government would need to work
closely with compute providers and the AI industry to monitor how AI
development changes in response to the introduction of such schemes and
develop responses.

An information-sharing mechanism could be a key tool that compute
providers could use to work together to identify actors purchasing
disaggregated compute from different companies. However,
information-sharing between competitors may be restricted by US
Antitrust Laws.\autocite{us_department_of_justice_antitrust_2023} Close engagement with legal and regulatory experts will be
required to carefully design an appropriate scheme, and/or statutory
protection could be achieved through legislation. Previous case studies
offer some hope: statements from the Department of Justice and Federal
Trade Commission noted that sharing information regarding cybersecurity
threats was appropriate and a well designed cyber threat sharing scheme
would not be likely to raise antitrust concerns.\autocite{federal_trade_commission_ftc_2014} Privacy preserving techniques, such as
Private Set Intersection computation,\autocite{nist_computer_security_research_center_brief_2021} can also be employed
effectively in support of information sharing.

Another detection evasion risk could arise from the use of shell
companies to obscure an entity's ultimate owners. The beneficial
ownership information reporting requirement, commencing January 2024 in
the US and is currently being adopted more broadly by FATF members,
could help decrease this risk by requiring companies to disclose
information on the people who ultimately own them.\autocite{financial_crimes_enforcement_network_fincen_2022} However, given the
strategic and economic importance of advanced AI, it is likely that
{[}malicious{]} actors will continue to try to obfuscate their
identities. Given the relatively small numbers of entities seeking to
access significant amounts of advanced AI compute in the near term, a
government enforcement team could consider undertaking their own
investigations and spot checks on companies.

\subsubsection{Enforcing the KYC
Scheme}\label{enforcing-the-kyc-scheme}

A key challenge for the scheme is how regulators can enforce compliance
with KYC requirements amongst compute providers. For example, how would
the regulator identify a compute provider's failure to effectively
implement the scheme, and/or deliberate non-compliance? Information and
expertise asymmetry between technologically informed compute providers
and nascent government regulation capability will add to this challenge.
Microsoft's advocacy for the introduction of KYC means there is some
industry goodwill to leverage, but the government will also have to
incorporate external expertise to guard against regulatory capture. As a
starting point, enlisting support from jurisdictions with experience in
AI regulation, such as the EU, as well as engaging AI researchers, could
help strengthen enforcement abilities. The administration of this scheme
would build a deeper understanding of AI in government, and enable more
effective AI policy more broadly.

Unlike in the parallel financial-sector crime of money laundering, the
AI sector does not have established entities and resources for
identifying and policing cloud compute access. However, following the
model of the financial sector, government could consider establishing a
whistleblower scheme with a financial reward to incentivize employees to
report suspected violations.

The scheme should be established to allow for flexible enforcement
tools, including the ability to direct a compute provider to pause or
cease providing compute to any entity found to be in violation of the
scheme. Enforcement should also be paired with the effective use of the
Bureau of Industry and Security's (BIS) Export Administration
Regulations. Where cloud compute is identified as supporting military,
intelligence, or security end uses that conflict with US interests, BIS
could then introduce export controls to prevent US compute providers
from providing this service.\autocite{dohmen_controlling_2023} The KYC scheme should also be
enforced through the application of penalties. To avoid a culture of
non-compliance like that seen in the early days of the Banking Secrecy
Act, regulators should demonstrate a willingness to issue non-trivial
penalties for deliberate non-compliance, in addition to playing a
regulatory coaching role. Depending on how the threat context evolves,
there may also be value in dedicating government resources to actively
seek out and investigate potential violations.\footnote{This could
  involve scouring the darkweb for evidence of undisclosed/underhand
  resale of substantial AI compute power; and/or intelligence gathering
  and analysis, depending on the nature of the identified risks.}

\subsection{Establishing government capacity for engagement,
regulation, and
enforcement}\label{establishing-government-capacity-for-engagement-regulation-and-enforcement}

To co-design, implement, administer, update, and enforce a KYC scheme
for advanced AI cloud compute, the US government should establish a new
unit in the Department of Commerce that is responsible for AI regulatory
policy. This unit should collaborate with and leverage expertise from
other relevant entities including the BIS, the National Institute of
Standards and Technology, FinCEN and AI industry and researchers. It
should also work closely with counterparts at the Department of Defense,
to ensure that national security needs are met, as well as the
Department of State, to help promulgate the scheme internationally.

To keep the scheme up-to-date and geared towards key risks, the
introduction of a KYC scheme should be accompanied by a process for
regular evaluation and iteration. Updates should be informed by
consultation with relevant agencies, stakeholder industries and
international partners. The US National AI Advisory Council (NAIAC),
established in 2022, could play a key role in providing cross-sectoral
advice in this process. While the NAIAC has a broad remit that includes
workforce, ethical, and leadership issues, it has a key role in ensuring
ongoing US AI leadership, coordination across civil and security
agencies, and matters relating to oversight of AI systems.\autocite{coggins_national_2022} To ensure that
technical expertise continues to inform updates, the NAIAC could
establish a working group on AI compute controls that brings together
private, public, and academic sector expertise.

\subsection{Engaging and cooperating with international
partners}\label{engaging-and-cooperating-with-international-partners}

While US dominance of cloud compute will render even domestic KYC
globally significant, international support for the scheme will help
maximize its effectiveness, particularly in the longer term. As the
leader in cloud service provision, the US is uniquely positioned to
shape global regulations and standards for advanced AI cloud compute.
There are an estimated 335 to 1325 data centers globally with a capacity
of above 10MW -- enough to enable a large AI training run -- with the
majority owned by US technology leaders.\autocite{pilz_compute_2023} However, only a minority of them actually host
the AI-specific compute and the exact numbers are unknown; KYC could
assist in developing a clearer picture. Nevertheless, unilateral US
action will be limited in shaping the market long term. Operating alone
could give rise to adverse outcomes, diminishing the attractiveness of
US compute providers and pushing those with greater privacy premiums to
lower regulatory environments. Over time, this could degrade US
leadership in compute and AI.

The US should therefore work with key international partners on an
aligned KYC scheme for advanced AI cloud compute. While nations' and
companies' adherence to the scheme could ultimately be enforced via the
threat of withholding US chip hardware exports, such a scheme will be
most effective if supported by goodwill and shared purpose. The US
should therefore engage first diplomatically, with a particular focus on
countries with significant data center architecture. These would include
European countries, which are estimated to host approximately 25 percent
of all data centers globally,\autocite{pilz_compute_2023} and Japan with
its world-leading supercomputer capabilities.\autocite{noauthor_june_nodate} Engagement with a broader set of likeminded
countries will also help to add momentum to the international
initiative. For information on these countries' positions on AI
regulation, refer to Box 5.

In addition to direct outreach on the shared risks, the US should
leverage partnerships with cross-border companies that can internally
advocate for a KYC regime. For example, because Microsoft has already
proposed implementing a KYC model for advanced AI,\autocite{smith_how_2023} it could help engage partners like the UK
that are seeking to prioritize innovation and light-touch regulation.

Following engagement with key partners, the US should work to establish
longer-term international architecture. Like the FATF in the financial
sector, an intergovernmental organization for AI compute controls could
help share risk information and align standards and best practices.
Given a focus on tactical and implementation matters, this organization
would be separate and complementary to existing international AI risk
groupings, like the OECD AI Policy Observatory.\autocite{wyckoff_globalpolicyai_2021}

\begin{tcolorbox}[breakable,boxrule=1pt,enhanced jigsaw, sharp corners,pad at break*=1mm,colbacktitle=gray!05,colback=gray!05,colframe=black,coltitle=black,toptitle=1mm,bottomtitle=1mm,fonttitle=\bfseries\centering,parbox=false,title=Box 5: Current positions on AI regulation and implications for
US KYC engagement]
\textbf{Europe}

The European Union's (EU) approaches to AI regulation will be shaped by
the EU AI Act. On 14 June 2023, the draft text of the law was agreed
upon between the European Parliament and the EU's executive branch, with
negotiations beginning with EU countries on the final form of the
law.\footnotemark\ Amongst other
measures, the draft law prohibits some specific AI uses (like real-time
facial recognition), introduces safety requirements for high-risk use
cases, and imposes risk assessments for foundation models.\footnotemark\ The draft law does not currently introduce specific risk
requirements based on level of compute used.\footnotemark\ Given the extensive negotiation and law-making
process involved in the EU AI Act, the EU and its members may be
hesitant to enter further discussions on additional restrictions. Yet
the passage of the AI Act would signify a focus on safety and security
that could be further enhanced through cloud KYC. Given the significant
regulatory burden of the EU AI Act, emphasizing the minimal compliance
costs of KYC will be key. The EU member state of the Netherlands has
also shown a willingness to align with US efforts to curb dangerous AI
proliferation to China.\footnotemark\

\textbf{Japan}

Like the Netherlands, Japan has already engaged closely with the US on
efforts to curb China's access to semiconductor technologies, which
impact AI development.\footnotemark\ However,
Japan's AI policies to date have focused on maximizing AI's positive
impact, rather than concentrating on risks. Japan considers further
regulation not yet necessary.\footnotemark\ Japan may
therefore be more motivated by the national security benefits of KYC,
rather than broader public safety concerns.

\textbf{UK}

The UK's 2023 policy paper \emph{A pro-innovation approach to AI
regulation} cautions against `placing unnecessary regulatory burdens on
those deploying AI,' and instead focuses on the context in which AI is
deployed.\footnotemark\ Nevertheless,
the UK is actively considering the role of measuring compute in the
governance of foundation models,\footnotemark\ and is a close strategic partner of the US. Given its
focus on making innovation easier and strengthening its leadership in
AI, the UK will be particularly sensitive to regulatory impacts of a KYC
scheme.

\textbf{Canada}

On 16 June 2022, Canada introduced the Artificial Intelligence and Data
Act into Parliament.\footnotemark\ The Act seeks to ensure that high
impact AI systems meet safety and human rights standards, to enable
enforcement and policy to keep up with technology developments, and to
prohibit reckless and malicious uses of AI.\footnotemark\ One of the principles for managing high-risk systems is
human oversight and monitoring, including in how systems are designed
and trained.\footnotemark\ KYC could be used to
support this effort.

\textbf{Australia}

Given its relatively small market size, Australia is aware that its
`ability to take advantage of AI supplied globally and support the
growth of AI in Australia will be impacted by the extent to which
Australia's responses are consistent with responses
overseas.'\footnotemark\ From May to
July 2023, the Australian government undertook a public consultation
process on how the government could support safe and responsible AI.
\end{tcolorbox}
\addtocounter{footnote}{-11}
\footnotetext{\cite{european_parliament_eu_2023}.}\stepcounter{footnote}
\footnotetext{\cite{european_parliament_eu_2023}.}\stepcounter{footnote}
\footnotetext{\cite{perrigo_eu_2023}.}\stepcounter{footnote}
\footnotetext{\cite{kharpal_netherlands_2023}.}\stepcounter{footnote}
\footnotetext{\cite{cash_china_2023}.}\stepcounter{footnote}
\footnotetext{\cite{habuka_japans_2023}.}\stepcounter{footnote}
\footnotetext{\cite{uk_office_for_artificial_intelligence_policy_2023}.}\stepcounter{footnote}
\footnotetext{\cite{uk_office_for_artificial_intelligence_policy_2023}.}\stepcounter{footnote}
\footnotetext{\cite{minister_of_innovation_science_and_industry_act_nodate}.}\stepcounter{footnote}
\footnotetext{\cite{government_of_canada_artificial_2023}.}\stepcounter{footnote}
\footnotetext{\cite{government_of_canada_artificial_2023}.}\stepcounter{footnote}
\footnotetext{\cite{australian_government_department_of_industry_science_and_resources_safe_2023}.}\stepcounter{footnote}

\section{Recommendations}\label{recommendations}

\subsection{The US Department of Commerce should work with
compute providers, industry stakeholders and AI researchers to develop a
domestic KYC scheme for advanced AI cloud
compute.}\label{the-us-department-of-commerce-should-work-with-compute-providers-industry-stakeholders-and-ai-researchers-to-develop-a-domestic-kyc-scheme-for-advanced-ai-cloud-compute.}

This should include:\vspace{-3pt}
\begin{itemize}
\item defining the threshold of AI compute at which the scheme would apply
  that captures frontier AI models and risks of bolstering an
  adversarial military's capability.
\item introducing requirements for compute providers above that threshold
  to verify customers' identities, keep records, report to government any
    entities that match government-specified `high-risk' profiles, and
    implement controls.
\item
  establishing a government unit for implementation, monitoring, and
  enforcement.
\item
  a clear process for evaluation and updates to the scheme.
\end{itemize}

\subsection{The Department of Commerce should update rules affecting the Export Administration Regulations to explicitly prohibit the provision of above-threshold
compute to  entities on the Entity List without a license.}

This would help prevent US resources from being used by entities acting contrary to US national security interests, while still enabling US industry leadership. The KYC scheme would provide the mechanism for US companies to effectively implement these controls.

\subsection{The Department of Commerce should implement this
scheme under the authority of a Presidential Executive Order,
potentially leveraging existing
efforts.}\label{the-department-of-commerce-should-implement-this-scheme-under-the-authority-of-a-presidential-executive-order-potentially-leveraging-existing-efforts.}

President Trump's 2021 Executive Order \emph{Taking Additional Steps to
Address the National Emergency with Respect to Significant Malicious
Cyber-Enabled Activities} charged the Department of Commerce with
introducing an obligation for providers of US IaaS to verify and record
the identity of persons applying for accounts in order to combat
malicious cyber activity.\autocite{the_white_house_executive_nodate} More recently,
the 2023 National Cybersecurity Strategy commits the Administration to
implementing this Executive Order.\autocite{the_white_house_national_2023} While the scope of this measure is broader
-- applying to all IaaS -- and its exact form unclear, there is an
opportunity to combine these schemes to create a tiered approach. Light
touch KYC could be implemented across the board, with comprehensive due
diligence obligations applied to the frontier model and strategic risk
threshold. Should the US Administration consider that this existing
authority is insufficient, the President should consider issuing a new
Executive Order explicitly granting authority for the KYC Scheme. This
additional authority would likely be needed to broaden the scheme to
capture non-CSP entities, such as compute hardware providers and
in-house use of compute for AI development.

\subsection{The Department of Commerce should work with the
Department of Justice and the Federal Trade Commission to publish
guidance on US Antitrust Laws in this
space.}\label{the-department-of-commerce-should-work-with-the-department-of-justice-and-the-federal-trade-commission-to-publish-guidance-on-us-antitrust-laws-in-this-space.}

This will help inform and enable a privacy-preserving
information-sharing mechanism to identify and address evasion efforts,
while ensuring that competition is not reduced and consumer protections
are upheld.

\subsection{The Department of State should work closely with the
Department of Commerce to build international support for an advanced AI
cloud compute KYC
scheme.}\label{the-department-of-state-should-work-closely-with-the-department-of-commerce-to-build-international-support-for-an-advanced-ai-cloud-compute-kyc-scheme.}

This should include partnering with entities with significant data
center capability and developing new international architecture to
support the scheme.

\section{Conclusion}

Amidst calls for more active management of AI, a KYC scheme will address a gap in export controls, and provide the US government with the
groundwork for greater oversight and control over risky AI developments.
While the scheme will not address all AI risks, it will allow for a more
precise and flexible approach to controlling risks from frontier AI
models and unwanted AI proliferation. It will also increase government
visibility over AI and allow for more proactive management. When scaled
internationally, this scheme has the potential to support global AI
governance architecture. Essential to the scheme's success is genuine
co-design with industry and experts. This partnership will be keen to
minimize harm while maximizing the benefits of AI.

\newpage
\section*{Appendix}
\appendix

\textbf{Practical Considerations for Implementing KYC based on
Compute
Threshold}

While compute serves as a crucial instrument in the proposed Know Your
Customer (KYC) scheme, it is important to acknowledge its limitations
and the associated open questions that might influence its effective
implementation.

\textbf{First, the total amount of compute utilized across various
models and tasks is inherently dynamic and not pre-defined.} The
complete aggregate compute, accumulating the computational power that a
system consumes over time, ultimately results in a total FLOP count,
becoming evident as it is being used. This dynamic accumulation implies
that identifying which customer surpasses the threshold preemptively is
challenging.\footnote{Furthermore, it's an oversimplification to assume
  that all the rented compute contributes solely to the system's
  training. The development process usually involves iterative
  refinements, so that the total compute usage could be two to four
  times the actual training compute.}

However, the practical scarcity of GPUs necessitates advance requests
for access, especially considering the substantial number required to
meet the discussed thresholds (thousands of GPUs). Developers also
prefer a large, interconnected cluster of GPUs rather than disparate
units distributed across different clusters or data centers, as this
facilitates high-bandwidth connections essential for AI development and
deployment. Moreover, the level of compute under consideration,
amounting to two-digit dollar millions, usually requires negotiations
with sales teams, rather than on-demand purchases (where CSPs might
already undertake something similar to a KYC to make sure that they are
being paid). Therefore, we suggest that, practically speaking, CSPs are
already aware of large compute orders and can then enact the required
KYC.

In scenarios where exact compute needs are indeterminate, continuous
monitoring of chip-hour accumulation is important. Warnings should be
set at varying levels, such as 50\% and 80\% of the threshold, to
maintain awareness of the customer when approaching limits. Upon meeting
the threshold, KYC procedures should be instigated, and if not
previously completed, access should be terminated.

Given the inherent characteristics of computational requirements in AI,
we anticipate a bimodal distribution where most actors either utilize
minimal amounts below the threshold or significantly larger quantities.
In most instances, users will either surpass the threshold from the
onset or they won't come remotely close to reaching it. This expectation
is attributed to the order of magnitude differences in AI compute usage
for AI systems, the connected monetary costs, and the resulting small
number of frontier AI developers. Consequently, the application of an
indicator value to establish a lower boundary, or warning level, proves
to be particularly effective in this context. It ensures that the
systems predominantly captured are those that either exceed or are in
close proximity to the set threshold.

\textbf{Second, the limitations in discerning specific insights about
user (AI) activities}, such as the phase of the AI lifecycle or model
architecture, are important considerations for the implementation of
this scheme. It's critical to leverage instruments that preserve the
privacy and integrity of user workloads. Despite these limitations, we
maintain optimism that the available tools and metrics are sufficiently
indicative and could be expanded by instigating further inquiries and
assurances from the customer without compromising their privacy.

Compute providers can easily access data related to total compute usage,
such as the number of chip hours and the type of chip, as they are
required for billing. However, the intrinsic privacy and confidentiality
clauses of cloud business prohibit providers from gaining insights into
the precise activities of the customer at the system level. This
restriction necessitates reliance on high-level, abstract metrics, which
are the only permissible indicators available for monitoring.

Our approach leverages available high-level metrics, which serve as
preliminary indicators prompting further checks, without delving into
more detailed aspects of user workloads. Potential concepts like proof
of training or the yet-to-be-explored proof of inference could allow
customers to validate their compute usage without requiring providers to
access more information than total compute usage.

In future work, we plan to investigate how various cluster-level
metrics, extending beyond chip-hours, can be leveraged by cloud
providers to facilitate enhanced oversight while preserving user
privacy. For instance, exploring metrics like network traffic patterns
could be instrumental in differentiating between AI training and
inference processes. We also urge the wider research community to
investigate these issues.

In this paper, we are primarily addressing the governance of AI systems
development. Questions about potential large-scale deployment, i.e.
inference compute, are outside the scope of this particular discussion.
However, the oversight of deployment compute might also be a future
policy tool for regulating AI deployment (Appendix A of \cite{obrien2023};
Brundage et al., forthcoming).

The limitations in acquiring detailed insights necessitate a balanced
approach, focusing on high-level indications without compromising user
privacy. The available metrics, which extend beyond type of chip and
chip-hours, are considered adequate for initial insights and prompt
actions. Further exploration of these topics is necessary, and we
advocate for continued research to refine and advance the methodologies
for achieving effective oversight while maintaining user trust and
privacy.

\section*{Acknowledgements}\label{acknowledgements}

We are grateful for valuable feedback and suggestion from Markus
Anderljung, Tim Fist, Konstantin Pilz, Alan Chan, and the rest of the
GovAI team. All errors are our own.

\newpage
\addcontentsline{toc}{section}{References}
\printbibliography

\end{document}